\documentclass[conference, final, letterpaper]{IEEEtran}

\usepackage{hyperref}
\usepackage{ifpdf}
\usepackage{cite}
\ifpdf
    \usepackage[pdftex]{graphicx}
    \graphicspath{{./fig/}}
    \DeclareGraphicsExtensions{.pdf}
\else
    \usepackage[dvips]{graphicx}
    \graphicspath{{./fig/}}
    \DeclareGraphicsExtensions{.eps}
\fi
\usepackage{theorem}
\usepackage[cmex10]{amsmath}
\usepackage{amssymb}
\usepackage{stmaryrd}
\usepackage[ruled,vlined,titlenumbered]{algorithm2e}

\newcommand{\erasure}{\ensuremath{\vartimes}}
\newcommand{\dmin}{\ensuremath{d_\mathrm{min}}}
\newcommand{\Hd}[1]{\ensuremath{\mathrm{d}_\mathrm{H}(#1)}}
\newcommand{\F}{\ensuremath{\mathbb{F}}}
\newcommand{\N}{\ensuremath{\mathbb{N}}}
\newcommand{\R}{\ensuremath{\mathbb{R}}}
\renewcommand{\vec}[1]{\ensuremath{\mathbf{#1}}}

\newcommand{\Ps}{\ensuremath{\mathrm{P}_{\sigma}}}
\newcommand{\Pst}{\ensuremath{\widetilde{\mathrm{P}}_{\sigma}}}

\newcommand{\estimate}[1]{\tilde{#1}}
\newcommand{\C}{\ensuremath{\mathcal{C}}}

{\theoremstyle{plain}
   \newtheorem{theorem}{Theorem}}

{\theoremstyle{plain}
   \newtheorem{problem}{Problem}}

{\theoremstyle{plain}
   \theorembodyfont{\mdseries}
   \newtheorem{example}{Example}}

\begin{document}

\title{Adaptive Single--Trial Error/Erasure Decoding for Binary Codes}

\IEEEoverridecommandlockouts

\author{
\authorblockN{Christian Senger, Vladimir R. Sidorenko, Steffen Schober, Martin Bossert}\thanks{This work has been supported by DFG,
Germany, under grants BO~867/17, BO~867/21-1, and RUS~436/113/0. Vladimir Sidorenko is on leave from IITP, Russian Academy of Sciences, Moscow, Russia.}
\authorblockA{\small Inst. of Telecommunications and Applied Information Theory\\
Ulm University, Ulm, Germany \\
\{christian.senger$\;\vert\;$vladimir.sidorenko$\;\vert\;$steffen.schober$\;\vert\;$martin.bossert\}@uni-ulm.de}
\and
\authorblockN{Victor V. Zyablov}
\authorblockA{\small Inst. for Information Transmission Problems\\
Russian Academy of Sciences, Moscow, Russia \\
zyablov@iitp.ru}
}

\maketitle

\begin{abstract}
We investigate adaptive single--trial error/erasure decoding of binary codes whose decoder is able to correct $\varepsilon$ errors and $\tau$ erasures if $\lambda\varepsilon+\tau\leq\dmin-1$. Thereby, $\dmin$ is the minimum Hamming distance of the code and $1<\lambda\leq 2$ is the tradeoff parameter between errors and erasures. The error/erasure decoder allows to exploit soft information by treating a set of most unreliable received symbols as erasures. The obvious question here is, how this erasing should be performed, i.e. how the unreliable symbols which must be erased to obtain the smallest possible residual codeword error probability are determined. This was answered before \cite{senger_sidorenko_zyablov:2009b} for the case of fixed erasing, where only the channel state and not the individual symbol reliabilities are taken into consideration. In this paper, we address the adaptive case, where the optimal erasing strategy is determined for every given received vector.
\end{abstract}

\section{Introduction}\label{sec:intro}

The idea of exploiting soft information from the transmission channel using hard--decision algebraic error/erasure decoders dates back to Forney \cite{forney:1966b, forney:1966a}. His {\em Generalized Minimum Distance (GMD)} decoding scheme applies a {\em Bounded Minimum Distance (BMD)} error/erasure decoder repeatedly, each time with a different number of erased most unreliable received symbols. Forney proved that the residual codeword error probability of GMD decoding approaches that of {\em Maximum Likelihood (ML)} decoding if the channel is good and the number of decoding trials is $\frac{\dmin}{2}$, where $\dmin$ is the minimum {\em Hamming distance} of the code. This explains why GMD decoding is frequently applied for concatenated coding schemes. There, the inner code is responsible for correcting a considerable amount of transmission channel errors. Thus, the input symbols for the outer decoder can be viewed as being transmitted over a {\em super channel}, which is composed of the transmission channel and the inner decoder. This super channel is always good if the parameters of the inner code are chosen appropriately.

The fundamental task of GMD decoding with given number of decoding trials is to find an erasing strategy which either maximizes the guaranteed decoding radius or minimizes the residual codeword error probability. Both measures can be optimized either in a fixed manner or adaptively. For fixed erasing, the erasing strategy depends only on the state of the transmission channel and remains unchanged for each received vector. The fixed approach essentially optimizes the overall worst--case measure. Adaptive erasing on the other hand takes every single received vector into consideration, choosing the optimal erasing strategy for exactly this specific received vector. Obviously, one can expect the adaptive approach to yield better results than the fixed approach, especially for mediocre channel conditions.

Different settings for optimal fixed erasing have been considered in \cite{senger_sidorenko_bossert_zyablov:2008a, senger_sidorenko_bossert_zyablov:2010a, sidorenko_senger_bossert_zyablov:2010} (radius maximization), \cite{senger_sidorenko_zyablov:2009b, senger_sidorenko_bossert_zyablov:2010b} (error probability minimization), and \cite{blokh_zyablov:1982, weber_abdel-ghaffar:2003} (both). Results about adaptive erasing can be found in \cite{kovalev:1986, sidorenko_senger_bossert_zyablov:2008, sidorenko_chaaban_senger_bossert:2009, chaaban_sidorenko_senger:2010, sidorenko_senger_bossert_zyablov:2010} (radius maximization).

In the present paper, we tackle the previously unconsidered problem of adaptive erasing with the target of minimizing the residual codeword error probability. In doing so, we restrict ourselves to one single decoding trial. This restriction allows to focus on the core of the problem and will be relaxed in future work. Furthermore, we assume binary antipodal signaling and a memoryless channel with soft output. The {\em Additive White Gaussian Noise (AWGN)} channel will serve as our main example for such channels.

The paper is organized as follows. In Section~\ref{sec:ee}, we briefly describe error/erasure decoding and introduce some required notations. In Section~\ref{sec:adaptive}, we derive an adaptive erasing strategy which minimizes the residual codeword error probability. In doing so, we apply basic techniques from probability theory like discrete random variables and probability generating functions. A computationally more efficient version of the erasing strategy is given in Section~\ref{sec:approx}. Simulation results are given in Section~\ref{sec:sim}, conclusions and an outlook to further research in Section~\ref{sec:conc}.

\section{Error/Erasure Decoding}\label{sec:ee}

We consider a binary code $\C(2; n, k, \dmin)$ with length $n$, dimension $k$ and minimum Hamming distance $\dmin$. For $\C$, we have a {\em $\lambda$--extended Bounded Distance} error/erasure decoder or simply {\em $\lambda$--decoder} $\mathrm{dec}_\C(\cdot)$ which is able to correct $\varepsilon$ errors and $\tau$ erasures if $\lambda\,\varepsilon+\tau\leq \dmin-1$. Here, $1<\lambda\leq 2$ is the tradeoff parameter between errors and erasures. For $\lambda=2$, the decoder is a traditional BMD error/erasure decoder. For {\em Bose--Chaudhuri--Hocquenghem (BCH)} codes, such error/erasure decoders are described e.g. in \cite{forney:1965, blahut:2003}.

At the transmitter, an information vector $\vec{a}\in\F_2^k$ is encoded into a codeword $\vec{c}\in\C\subseteq\F_2^n$. The binary symbols $c_i$, \mbox{$i=0, \ldots, n-1$}, are then mapped to binary antipodal signals $x_i:=-1^{c_i}\in\{-1, +1\}$, which are transmitted over the channel. Each transmitted symbol $x_i$ is distorted by the channel to a received symbol $y_i\in\R$. The $\lambda$--decoder can only handle hard input, hence the real received symbols must be mapped to symbols of the binary field $\F_2$. This can be accomplished by the {\em Heaviside--like function}
\begin{equation*}
  \alpha:=\left\{\begin{array}{rll}
                    \R & \longrightarrow & \R\\
                    y & \longmapsto &
                      \left\{\begin{array}{ll}
                      -1, & \text{if}\;y\leq0\\
                      +1, & \text{if}\;y> 0
                      \end{array}\right.
                    \end{array}\right.,
\end{equation*}
which essentially extracts the sign of a real received symbol, and the inverse mapping function
\begin{equation*}
  \beta:=\left\{\begin{array}{rll}
                    \{-1, +1\} & \longrightarrow & \F_2\\
                    y & \longmapsto &
                      \left\{\begin{array}{ll}
                      1, & \text{if}\;y=-1\\
                      0, & \text{if}\;y=+1
                      \end{array}\right.
                    \end{array}\right.,
\end{equation*}
which maps real symbols to symbols of $\F_2$. The binary vector
\begin{equation}\label{eqn:recvd}
\vec{r}:=(\beta(\alpha(y_0)), \ldots, \beta(\alpha(y_{n-1})))
\end{equation}
is a distorted version of the transmitted codeword $\vec{c}$ and could be fed into the $\lambda$--decoder for traditional errors--only decoding. Decoding would be successful if for the number $\varepsilon:=\Hd{\vec{c}, \vec{r}}$ of errors in $\vec{r}$ holds $\lambda\,\varepsilon\leq \dmin-1$ or, in more familiar notation, $\varepsilon\leq \left\lfloor\frac{\dmin-1}{\lambda}\right\rfloor$. Here, $\Hd{\cdot, \cdot}$ is the Hamming distance between two vectors of equal length.

Let $\Ps(\cdot|\cdot)$ be the transition probability of the memoryless channel, the parameter $\sigma$ marks the channel state. Then, using {\em Bayes' Theorem}, we can calculate for each received symbol $y$ the probability $h_\sigma(y)$ that $-\alpha(y)$ was transmitted, i.e. a transmission error occurred.
\begin{align*}
  h_\sigma(y)  & :=   \Ps(-\alpha(y)\,|\,y)\nonumber\\
   & =  \frac{\Ps(y\,|-\alpha(y)) \Pr(-\alpha(y))}{\Pr(y)}\\
   & =  \frac{%
          \Ps(y\,|-\alpha(y)) \Pr(-\alpha(y))%
          }{%
          \Ps(y\,|\alpha(y)) \Pr(\alpha(y))+\Ps(y\,|-\alpha(y)) \Pr(-\alpha(y))%
          }\\
  & =  \frac{%
          \Ps(y\,|-\alpha(y))%
          }{%
          \Ps(y\,|\alpha(y))+\Ps(y\,|-\alpha(y))%
          },
\end{align*}
where the last equality follows from the reasonable assumption $\Pr(-\alpha(y))=\Pr(\alpha(y))=\frac{1}{2}$ of equiprobable transmitted symbols. It is justified to denote $h_\sigma(y)$ as {\em unreliability value} of the received symbol $y$. The greater $h_\sigma(y)$, the higher the probability that $y$ is an erroneous symbol. W.l.o.g. let us from now on assume that the symbols of the received vector $\vec{y}$ (and by (\ref{eqn:recvd}) also $\vec{r}$) are ordered according to their unreliability value, i.e. $h_\sigma(y_0)\geq \cdots\geq h_\sigma(y_{n-1})$.

We obtain a new received vector by erasing the $\tau$ most unreliable symbols in $\vec{r}$. This new vector is denoted by
\begin{equation*}
  \vec{r}_\tau:=(\underbrace{\erasure, \ldots, \erasure}_{\tau\;\text{times}}, r_\tau, \ldots, r_{n-1}).
\end{equation*}
The $\lambda$--decoder is capable of decoding $\vec{r}_\tau$ as long as \mbox{$\lambda\,\varepsilon+\tau\leq \dmin-1$}, where $\varepsilon$ is the number of errors in the non--erased symbols $r_\tau, \ldots, r_{n-1}$. The number of erasures is the decoder's degree of freedom, so the task of an adaptive error/erasure decoder is as follows.

\begin{problem}\label{prob:1}
For given received vector $\vec{y}=(y_0, \ldots, y_{n-1})$ with ordered unreliabilities $h_\sigma(y_0)\geq \cdots\geq h_\sigma(y_{n-1})$ and channel state $\sigma$ find the optimal number $0\leq\tau_\sigma^\star\leq\dmin-1$ of erased most unreliable symbols such that the residual codeword error probability of decoding $\vec{r}_{\tau^\star}$ with the $\lambda$--decoder $\mathrm{dec}_\C(\cdot)$ is minimized.
\end{problem}

In the following section we provide an exact solution to Problem~\ref{prob:1} which is computationally expensive. In Section~\ref{sec:approx} we give a very good approximated solution which is computationally efficient.

\section{Derivation of an Adaptive Erasing Strategy}\label{sec:adaptive}

To solve Problem~\ref{prob:1} it is required to express the residual codeword error probability after adaptive error/erasure decoding as a function of the number $\tau$ of erased symbols. We accomplish this using basic techniques from probability theory.

Let the discrete random variables $X_i$, $i=0, \ldots, n-1$ be defined by
\begin{equation*}
  X_i:=\left\{\begin{array}{rl}
              1, & \text{if}\;y_i\;\text{is erroneous}\,(y_i\neq x_i)\\
              0, & \text{if}\;y_i\;\text{is correct}\,(y_i= x_i)
              \end{array}\right..
\end{equation*}
The probabilities of the two possible values of $X_i$ are determined by the unreliability value of symbol $y_i$, i.e. \mbox{$\Pr(X_i=1)=h_\sigma(y_i)$} and $\Pr(X_i=0)=1-h_\sigma(y_i)$.

Since $X_i$ takes on only nonnegative integer values, its {\em probability generating function (PGF)} \cite{feller:1968, gubner:2006} is given by
\begin{align}
  G_{\sigma, X_i}(\rho) & := \mathrm{E}\{\rho^{X_i}\}\label{eqn:def}\\
  & = \Pr(X_i=0)+\rho \Pr(X_i=1)\nonumber\\
  & = 1-h_\sigma(y_i)+\rho h_\sigma(y_i).\nonumber
\end{align}

Assume that the $\tau$ most unreliable symbols of $\vec{r}$ are erased and $\vec{r}_\tau$ is fed into the $\lambda$--decoder. Then, there are $\varepsilon$, \mbox{$0\leq \varepsilon\leq n-\tau$}, erroneous symbols among the non--erased $n-\tau$ symbols. We can model their number with a new random variable $Y_\tau$ using the random variables $X_i$, $i=\tau, \ldots, n-1$.
\begin{equation*}
  Y_\tau := \sum_{i=\tau}^{n-1} X_i.
\end{equation*}

We obtain
\begin{align}
  G_{\sigma,Y_\tau}(\rho) & := \mathrm{E}\{\rho^{Y_\tau}\}\nonumber\\
  & = \mathrm{E}\{\rho^{X_\tau+\cdots+X_{n-1}}\}\nonumber\\
  & = \mathrm{E}\{\rho^{X_\tau}\cdot\cdots\cdot\rho^{X_{n-1}}\}\label{eqn:prod}\\
  & = \mathrm{E}\{\rho^{X_\tau}\}\cdot\cdots\cdot\mathrm{E}\{\rho^{X_{n-1}}\}\nonumber\\
  & = \prod_{i=\tau}^{n-1} G_{\sigma, X_i}(\rho)\label{eqn:repl}
\end{align}
for the PGF of $Y_\tau$, i.e. the PGF of $Y_\tau$ is the product of the PGFs of the $X_\tau, \ldots, X_{n-1}$ and thereby known. Note that the expectation of the product in (\ref{eqn:prod}) can be written as a product of expectations since the channel is memoryless and thus the $X_i$ are independent. The product (\ref{eqn:repl}) results directly from the definition (\ref{eqn:def}) of the $G_{\sigma, X_i}$.

Using the PGF of $Y_\tau$ we can calculate the probability that there are $\varepsilon$, $0\leq\varepsilon\leq n-\tau$, errors in $\vec{r}_\tau$ by
\begin{equation}\label{eqn:probepsilon}
  \Pr(Y_\tau=\varepsilon):=\left.\frac{G_{\sigma, Y_\tau}^{(\varepsilon)}(\rho)}{\varepsilon!}\right|_{\rho=0},
\end{equation}
where the superscript $^{(\varepsilon)}$ denotes the $\varepsilon$-th derivative.

Recall that the $\lambda$--decoder is capable of decoding $\varepsilon$ errors and $\tau$ erasures if $\lambda\,\varepsilon+\tau\leq \dmin-1$. In case of $\tau$, \mbox{$0\leq\tau\leq\dmin-1$}, erasures the decoder will fail if the number of errors in the non--erased symbols is greater than $\frac{\dmin-1-\tau}{\lambda}$. Using (\ref{eqn:probepsilon}), the probability of this event is determined by
\begin{align}
  \Pr\left(Y_\tau>\frac{\dmin-1-\tau}{\lambda}\right) & =%
  1-\sum_{\varepsilon=0}^{\left\lfloor\frac{\dmin-1-\tau}{\lambda}\right\rfloor} \Pr(Y_\tau=\varepsilon)\nonumber\\
  & =: \Ps(\tau).\label{eqn:prob}
\end{align}
$\Ps(\tau)$ is the residual codeword error probability as a function of the channel state $\sigma$ and the number $\tau$ of erased symbols. Hence, the optimal choice of $\tau$ is
\begin{align}
  \tau_\sigma^\star & := \arg\min_{0\leq\tau\leq\dmin-1}\left\{%
    \Ps(\tau)
  \right\}\label{eqn:argmin}\\
  & = \arg\max_{0\leq\tau\leq\dmin-1}\left\{%
    \sum_{\varepsilon=0}^{\left\lfloor\frac{\dmin-1-\tau}{\lambda}\right\rfloor} \Pr(Y_\tau=\varepsilon)
  \right\}.\label{eqn:argmax}
\end{align}

The residual codeword error probability is minimized by erasing the $\tau_\sigma^\star$ most unreliable symbols since from (\ref{eqn:argmin}) we obtain
\begin{equation*}
  \Ps^\star:=\Ps(\tau_\sigma^\star)=\min_{0\leq\tau\leq\dmin-1}\left\{%
    \Ps(\tau)
  \right\},
\end{equation*}
which proves that adaptive erasing with $\tau_\sigma^\star$ as in (\ref{eqn:argmin}) is at least as good as errors--only decoding with $\tau=0$ and single--trial fixed erasing with some $\tau_{\mathrm{fixed}}^\star$, $0\leq\tau_{\mathrm{fixed}}^\star\leq\dmin-1$ in terms of the achievable residual codeword error probability.

Using the results from this section we can state Algorithm~\ref{alg:exact} for optimal adaptive error/erasure decoding. It provides an exact solution for Problem~\ref{prob:1}.

\begin{algorithm}[htbp]
\SetKwInOut{Input}{input}\SetKwInOut{Output}{output}
\dontprintsemicolon
\linesnumbered
{%

  \Input{$\C(2; n, k, \dmin)$, $\vec{y}\in\R^n$, $\sigma$, $\lambda$--decoder $\mathrm{dec}_\C(\cdot)$}

  calculate $h_\sigma(y_0), \ldots, h_\sigma(y_{n-1})$\;
  
  sort $\vec{y}$ s.t. $h_\sigma(y_0)\geq\cdots\geq h_\sigma(y_{n-1})$\tcp*[f]{$\mathcal{O}(n^2)$}\;
  
  $\vec{r}\leftarrow (\beta(\alpha(y_0)), \ldots, \beta(\alpha(y_{n-1})))$\;

  \For(\tcp*[f]{$\mathcal{O}(n^2)$}){$\tau=0, \ldots, \dmin-1$}{
    calculate $G_{\sigma, Y_\tau}(\rho)$\;
  }
     
  $m\leftarrow 1$\;
  
  \For(\tcp*[f]{$\mathcal{O}(n^2\dmin)$}){$\tau\leftarrow 0$ \KwTo $\dmin-1$}{
    \For(\tcp*[f]{$\mathcal{O}(n^2)$}){$\varepsilon=0, \ldots, \frac{\dmin-1-\tau}{\lambda}$}{
    calculate $G_{\sigma, Y_\tau}(\rho)^{(\varepsilon)}|_{\rho=0}$\;
    }
    \If(\tcp*[f]{$\mathcal{O}(n\dmin)$}){$\Ps(\tau)<m$}{
      $\tau_\sigma^\star\leftarrow\tau$\;
      $m\leftarrow \Ps(\tau)$\;
    }
  }

  calculate $\vec{r}_{\tau_\sigma^\star}$ from $\vec{r}$\;

  revoke sorting of $\vec{r}_{\tau_\sigma^\star}$\;

  \KwRet $\mathrm{dec}_\C\left(\vec{r}_{\tau_\sigma^\star}\right)$\tcp*[f]{$\mathcal{O}(n^2)$}

  \Output{codeword estimate $\estimate{\vec{c}}\in\C$ or erasure $\erasure$}

}
\caption{Optimal Adaptive Error/Erasure Decoding}
\label{alg:exact}
\end{algorithm}

The drawback of Algorithm~\ref{alg:exact} is its computational complexity. Sorting a vector of length $n$ in line~2 has complexity $\mathcal{O}(n^2)$ and can be accomplished in place e.g. by the {\em bubble sort} algorithm \cite{knuth:1998}. Calculating the PGFs $G_{\sigma, Y_\tau}(\rho)$, \mbox{$\tau=0, \ldots, \dmin-1$}, in lines~4--5 essentially means multiplying $n$ polynomials $G_{\sigma, X_i}(\rho)$, each with degree $1$. This can be done efficiently using $n$ {\em Fast Fourier Transforms (FFT)} of length $n$ and componentwise multiplication of the frequency domain coefficients. Since the input polynomials for the FFT have degree $1$ (e.g. only two non--zero coefficients), {\em $2$--pruned} FFTs \cite{sorensen_burrus:1993} with complexity $\mathcal{O}(n)$ can be used. The $n$ $2$--pruned FFTs together have complexity $\mathcal{O}(n^2)$ and the number of componentwise multiplications is $n^2$. The required single inverse FFT of length $n$ has complexity $\mathcal{O}(n\log(n))$. Hence, the complexity of lines~4--5 is $\mathcal{O}(n^2)$. The loop in lines~8--9 requires the evaluation of $\left\lfloor\frac{\dmin-1-\tau}{\lambda}\right\rfloor+1$ derivatives at $\rho=0$. This can be accomplished with complexity $\mathcal{O}\left(\left(\left\lfloor\frac{\dmin-1-\tau}{\lambda}\right\rfloor+1\right) n\right)\subseteq\mathcal{O}(n^2)$ using an algorithm from Pankiewiczs \cite{pankiewicz:1968} which is based on {\em Horner's Scheme}. The resulting values are required for the calculation of the $\Ps(\tau)$ in line~10 as in (\ref{eqn:prob}). For each $\Ps(\tau)$, a sum over $\left\lfloor\frac{\dmin-1-\tau}{\lambda}\right\rfloor+1$ probabilities $\Pr(Y_\tau=\varepsilon)$ has to be calculated. Using the pre--computed values from lines~8--9, this can be accomplished with complexity $\mathcal{O}(n\dmin)$. Since the loop in lines~7--12 is executed $\dmin$ times, its complexity is $\mathcal{O}(n^2\dmin)$. The complexity for $\lambda$--decoding in line~15 is $\mathcal{O}(n^2)$. Altogether the computational complexity of Algorithm~\ref{alg:exact} is $\mathcal{O}(n^2\dmin)\subseteq\mathcal{O}(n^3)$.

Section~\ref{sec:approx} addresses a computationally more efficient version of the algorithm which uses very good approximations of the $\Ps(\tau)$.

\begin{example}\label{ex:127}
We consider the BCH code $\C(2; 127, 36, 31)$ with a traditional BMD error/erasure decoder, i.e. $\lambda=2$. The symbols $\{-1, +1\}$ are transmitted over an AWGN channel. In this case, the unreliability of received symbol $y$ is
\begin{equation*}
  h_\sigma(y)=h_{\sigma, \mathrm{AWGN}}(y):=\frac{1}{1+\exp\left(\frac{2 y \alpha(y)}{\sigma^2}\right)}.
\end{equation*}
Throughout the paper $\exp(\cdot)$ and $\log(\cdot)$ have base $e$. We assume $\mathrm{SNR}=0\,\mathrm{dB}$, and obtain \mbox{$\sigma=\sqrt{\frac{1}{2}\cdot 10^\frac{-\mathrm{SNR}}{10}}=\sqrt{0.5}$}. Figure~\ref{fig:exact} depicts the operation of the loop in lines~7--12 of Algorithm~\ref{alg:exact}. For each $\tau=0, \ldots, 30$ and $\varepsilon=0, \ldots, \frac{30-\tau}{2}$ the probabilities $\Pr(Y_\tau=\varepsilon)$ are calculated. Each \mbox{$\Pr(Y_\tau=\varepsilon)$} is represented by one point in Figure~\ref{fig:exact}. This allows to calculate the sums in the maximization term of (\ref{eqn:argmax}). Each of the sums is the sum over one slice of the point surface in Figure~\ref{fig:exact} in $\varepsilon$--direction. The optimal choice of $\tau$ is the slice whose sum is maximal, in case of the considered codeword/transmission/received vector the optimization yields $\tau_{\sqrt{0.5}}^\star=4$.

\begin{figure}[htbp]
\centering
\includegraphics[width=252pt]{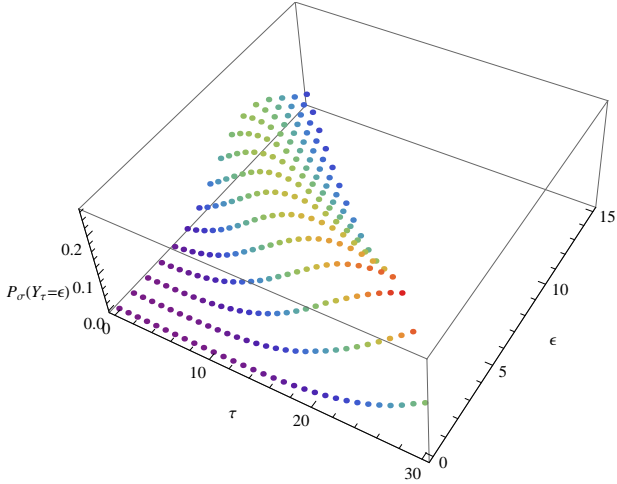}
\caption{Point surface consisting of the probabilities $\Pr(Y_\tau=\varepsilon)$, where $\tau=0, \ldots, 30$ and $\varepsilon=0, \ldots, \frac{30-\tau}{2}$.}
\label{fig:exact}
\end{figure}

\end{example}

\section{Computationally Efficient Adaptive Erasing}\label{sec:approx}

In this section, we present a technique which allows to reduce the computational complexity of Algorithm~\ref{alg:exact} from cubic in $n$ to $\mathcal{O}(n^2\sqrt[4]{n})$. It utilizes an approximation of the probabilities $\Ps(\tau)$, $\tau=0, \ldots, \dmin-1$. This approximation is based on the following result by Hoeffding \cite{hoeffding:1963}.

\begin{theorem}[Hoeffding Bound]\label{thm:hoeffding}
Let $A_0, \ldots, A_{m-1}$ be $m$ independent random variables with finite first and second moments, which are almost surely bounded, i.e.
\begin{equation*}
  \Pr(A_i-E\{A_i\}\in[a_i, b_i])=1,\quad i=0, \ldots, m-1,
\end{equation*}
where $E\{\cdot\}$ denotes the expectation of a random variable. Then, for the sum $S=A_0+\cdots+A_{m-1}$ and $t>0$ holds
\begin{equation*}
  \Pr(|S-E\{S\}|\geq mt)\leq 2\exp\left( -\frac{2 m^2 t^2}{\sum_{i=0}^{m-1} (b_i-a_i)^2}\right),
\end{equation*}
\end{theorem}

\begin{figure}[htbp]
\centering
\includegraphics[width=242pt]{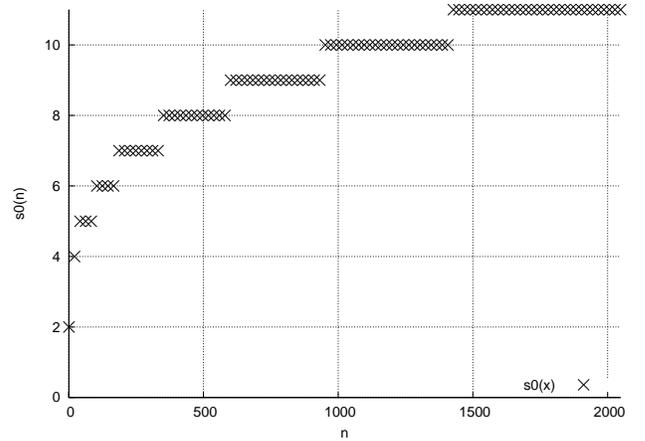}
\caption{Value of $s_0(n)$, $n=1, \ldots, 2048$, for precision goal $10^{-2}$.}
\label{fig:s0}
\end{figure}

We apply Theorem~\ref{thm:hoeffding} to $Y_\tau=\sum_{i=\tau}^{n-1} X_i$, i.e. $m=n-1-\tau$. By definition, we have $X_i\in\{0, 1\}$ and thus 
\begin{equation*}
\sum_{i=0}^{m-1} (b_i-a_i)^2=m=n-1-\tau.
\end{equation*}
We obtain
\begin{multline*}
  \Pr\left(|Y_\tau-E\{Y_\tau\}|\geq t(n-1-\tau)\right) \\
  \leq 2\exp\left(-2 t^2 (n-1-\tau)\right).
\end{multline*}
This means that the sum of the probabilities
\begin{multline}\label{eqn:neglectedprob}
\Pr(Y_\tau=0), \ldots, \Pr(Y_\tau=E\{Y_\tau\}-t), \\ \Pr(Y_\tau=E\{Y_\tau\}+t), \ldots, \Pr(Y_\tau=\dmin-1)
\end{multline}
 is exponentially decreasing with $t$. We can conclude that the sum in (\ref{eqn:prob}) is dominated by only a small set of probabilities in proximity to the expectation $E\{Y_\tau\}$. Let us set $t:=\frac{s}{n-1-\tau}$. We obtain
\begin{align*}
  \Pr(|Y_\tau-E\{Y_\tau\}|\geq s) &\leq 2 \exp\left( -\frac{2 s^2}{\sqrt{n-1-\tau}}\right)\\
  &\leq 2 \exp\left(-\frac{2 s^2}{\sqrt{n}}\right),
\end{align*}
i.e. the contribution of the probabilities from (\ref{eqn:neglectedprob}) in (\ref{eqn:prob}) is less than $2 \exp\left(-\frac{2 s^2}{\sqrt{n}}\right)$. This fact can also be observed in Figure~\ref{fig:exact}: The probabilities $\Pr(Y_\tau=\varepsilon)$ diminish quickly around the expectation of each slice in $\varepsilon$-direction.
To obtain a good approximation (with precision goal $10^{-2}$), let us select $s$ such that
\begin{align*}
  2 \exp\left(-\frac{2 s^2}{\sqrt{n}}\right) &< 10^{-2} \Longleftrightarrow\\
  s &> \sqrt{-\frac{\log(0.5\cdot 10^{-2})}{2} \sqrt{n}}.
\end{align*}
We define
\begin{equation*}
  s_0:=\left\{\begin{array}{rll}
                    \N\setminus\{0\} & \longrightarrow & \N\\
                    n & \longmapsto &
                    \left\lfloor\sqrt{-\frac{\log(0.5\cdot 10^{-2})}{2} \sqrt{n}}\right\rfloor+1
                    \end{array}\right..
\end{equation*}
Figure~\ref{fig:s0} shows the value of $s_0(n)$ for a practical range of code lengths $n$.

Eventually, the Hoeffding bound justifies to neglect
\begin{multline*}
\Pr(Y_\tau=0), \ldots, \Pr(Y_\tau=E\{Y_\tau\}-s_0(n)), \\ \Pr(Y_\tau=E\{Y_\tau\}+s_0(n)), \ldots, \Pr(Y_\tau=\dmin-1)
\end{multline*}
in the sum of (\ref{eqn:prob}). As a result, we obtain very good approximations for $\Ps(\tau)$ if we calculate the sum in (\ref{eqn:prob}) over at most $2s_0(n)$ elements, i.e.
\begin{equation*}
  \Ps(\tau) \approx \Pst(\tau):=
  1-\hspace{-0.25cm} \sum_{\varepsilon=\max\left\{\left\lceil E\{Y_\tau\}\right\rceil-s_0(n),\, 0\right\}}^{\left\lfloor\min\left\{E\{Y_\tau\}+s_0(n), \frac{\dmin-1-\tau}{\lambda}\right\}\right\rfloor} \hspace{-0.25cm}\Pr(Y_\tau=\varepsilon).
\end{equation*}
The required expectation can be easily calculated using the PGF (\ref{eqn:repl}) of $Y_\tau$, i.e.
\begin{equation}\label{eqn:expectation}
  E\{Y_\tau\}:=\left.G_{\sigma, Y_\tau}(\rho)^{(1)}\right|_{\rho=1},
\end{equation}
where the superscript $^{(1)}$ denotes the first derivative.

We use the previous results to state Algorithm~\ref{alg:approx} which solves Problem~\ref{prob:1} with high precision and better computational complexity than Algorithm~\ref{alg:exact}.

\begin{algorithm}[htbp]
\SetKwInOut{Input}{input}\SetKwInOut{Output}{output}
\dontprintsemicolon
\linesnumbered
{%

  \Input{$\C(2; n, k, \dmin)$, $\vec{y}\in\R^n$, $\sigma$, $s_0(n)$,\\$\lambda$--decoder $\mathrm{dec}_\C(\cdot)$}

  calculate $h_\sigma(y_0), \ldots, h_\sigma(y_{n-1})$\;
  
  sort $\vec{y}$ s.t. $h_\sigma(y_0)\geq\cdots\geq h_\sigma(y_{n-1})$\tcp*[f]{$\mathcal{O}(n^2)$}\;
  
  $\vec{r}\leftarrow (\beta(\alpha(y_0)), \ldots, \beta(\alpha(y_{n-1})))$\;

  \For(\tcp*[f]{$\mathcal{O}(n^2)$}){$\tau=0, \ldots, \dmin-1$}{
    calculate $G_{\sigma, Y_\tau}(\rho)$\;
  }
     
  $m\leftarrow 1$\;
  
  \For(\tcp*[f]{$\mathcal{O}(n\sqrt[4]{n}\dmin)$}){$\tau\leftarrow 0$ \KwTo $\dmin-1$}{
    calculate $E\{Y_\tau\}$\;
    $l\leftarrow \max\{\left\lceil E\{Y_\tau\}\right\rceil-s_0(n), 0\}$\;
    $u\leftarrow \left\lfloor\min\{E\{Y_\tau\}+s_0(n), \frac{\dmin-1-\tau}{\lambda}\}\right\rfloor$\;
    \For(\tcp*[f]{$\mathcal{O}(n\sqrt[4]{n})$}){$\varepsilon=l, \ldots, u$}{
    calculate $G_{\sigma, Y_\tau}(\rho)^{(\varepsilon)}|_{\rho=0}$\;
    }
    \If(\tcp*[f]{$\mathcal{O}(n\sqrt[4]{n})$}){$\Pst(\tau)<m$}{
      $\tau_\sigma^\star\leftarrow\tau$\;
      $m\leftarrow \Pst(\tau)$\;
    }
  }

  calculate $\vec{r}_{\tau_\sigma^\star}$ from $\vec{r}$\;

  revoke sorting of $\vec{r}_{\tau_\sigma^\star}$\;

  \KwRet $\mathrm{dec}_\C\left(\vec{r}_{\tau_\sigma^\star}\right)$\tcp*[f]{$\mathcal{O}(n^2)$}

  \Output{codeword estimate $\estimate{\vec{c}}\in\C$ or erasure $\erasure$}

}
\caption{Efficient Adaptive Error/Erasure Decoding}
\label{alg:approx}
\end{algorithm}

Algorithm~\ref{alg:approx} has some differences compared to Algorithm~\ref{alg:exact}, we will now briefly analyze their computational complexity.

Lines~1--6 remain unchanged, sorting, mapping to symbols of $\F_2$ and pre--calculation of the PGFs is the same for both the exact the the approximative algorithms. The loop in lines~7--15 starts with the calculation of the expectation $E\{Y_\tau\}$ according to (\ref{eqn:expectation}). This can be accomplished with linear complexity. In lines~9--10, lower and upper bounds for the loop in lines~11--12 are calculated, using essentially $E\{Y_\tau\}$ and the input parameter $s_0(n)$. Since $s_0(n)$ grows with $\sqrt[4]{n}$, the loop in lines~11--12 calculates the value of $\mathcal{O}(\sqrt[4]{n})$ subsequent derivatives of the PGF $G_{\sigma, Y_\tau}(\rho)$. The complexity of this calculation is $\mathcal{O}(n\sqrt[4]{n})$ using Pankiewiczs' algorithm \cite{pankiewicz:1968}. The calculation of $\Pst(\tau)$ in line~13 involves summation of $2\sqrt[4]{n}$ probabilities $\Pr(Y_\tau=\varepsilon)$. Using the pre--computed values of the derivatives from lines~11--12, each $\Pr(Y_\tau=\varepsilon)$ can be calculated with complexity linear in $n$, hence $\Pst(\tau)$ can be calculated with complexity $\mathcal{O}(n\sqrt[4]{n})$. Note that calculating $\Ps(\tau)$ in Algorithm~\ref{alg:exact} is in $\mathcal{O}(n\dmin)$. Alltogether, the loop in lines~7--15 is in $\mathcal{O}(n\sqrt[4]{n}\dmin)$ and thus the overall complexity of Algorithm~\ref{alg:approx} is $\mathcal{O}(n^2\sqrt[4]{n})$.

\section{Simulation Results}\label{sec:sim}

After the derivations of two adaptive error/erasure decoding algorithms in Sections~\ref{sec:adaptive} and \ref{sec:approx}, we devote this section to the analysis of their performance and behavior. First, we consider the short BCH code $\C(2; 31, 16, 7)$, a traditional BMD decoder with $\lambda=2$ and an AWGN channel in the range between $0\,\mathrm{dB}$ and $6\,\mathrm{dB}$.

\begin{figure}[htbp]
\centering
\makebox[0pt][l]{\includegraphics[width=252pt]{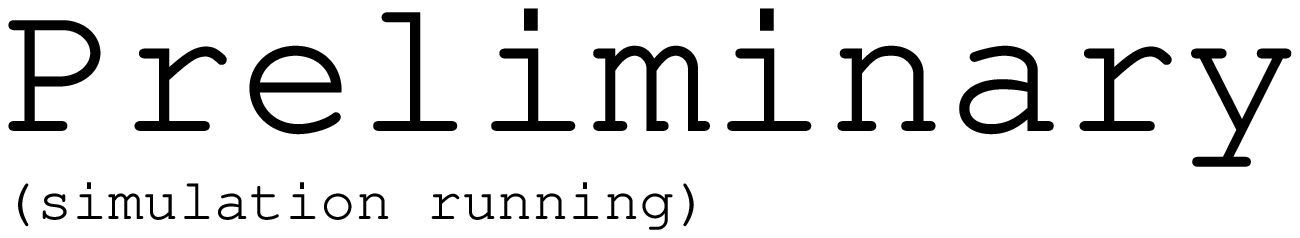}}%
\includegraphics[width=252pt]{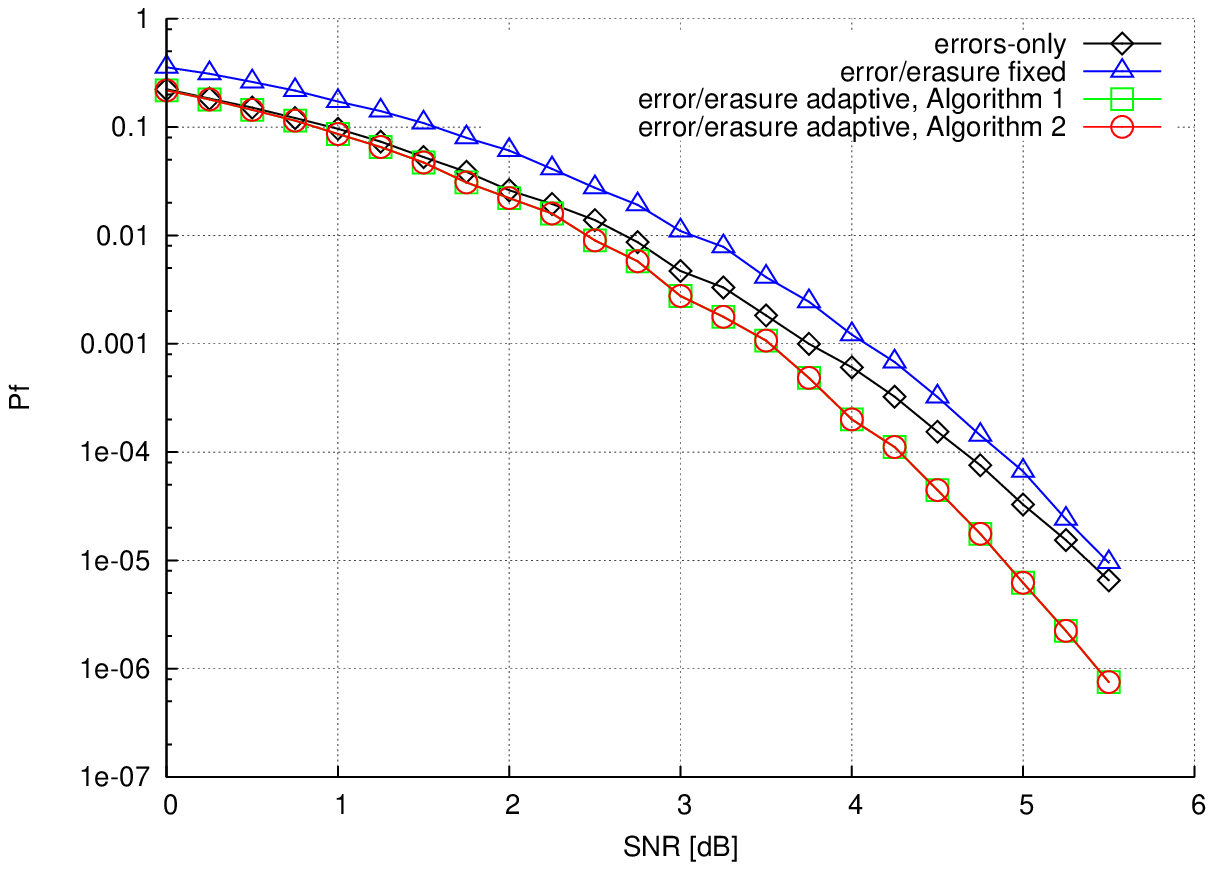}
\caption{Simulation results for $\C(2; 31, 16, 7)$.}
\label{fig:sim31}
\end{figure}

Figure~\ref{fig:sim31} shows the simulation results. The black curve (diamonds) denotes traditional errors--only decoding. The green curve (squares) shows the result of Algorithm~\ref{alg:exact}. It is not distinguishable from the red curve (circles) showing the result of the computationally more efficient Algorithm~\ref{alg:approx}. For reference, the figure also contains the result of error/erasure decoding with fixed erasing (blue curve, triangles) as in \cite{senger_sidorenko_zyablov:2009b}. The aforementioned result assumes very good channel conditions, hence its performance is bad in the considered range. However, there is a crossing point with the errors--only curve and we showed that the gain of optimal fixed erasing is $1.5\,\mathrm{dB}$ for an infinitively good channel. Note that the simulation confirms our observation from Section~\ref{sec:adaptive}, that Algorithm~\ref{alg:exact} must be as least as good as errors--only decoding {\bf and} error/erasure decoding with optimal fixed erasing.

For the second simulation, we reconsider the setting of Example~\ref{ex:127}, i.e. the BCH code $\C(2; 127, 36, 31)$. We observe that Algorithm~\ref{alg:approx} enables a reduction of the residual codeword error probability starting at around $\mathrm{SNR}=1\,\mathrm{dB}$.

\begin{figure}[htbp]
\centering
\makebox[0pt][l]{\includegraphics[width=252pt]{preliminary}}%
\includegraphics[width=252pt]{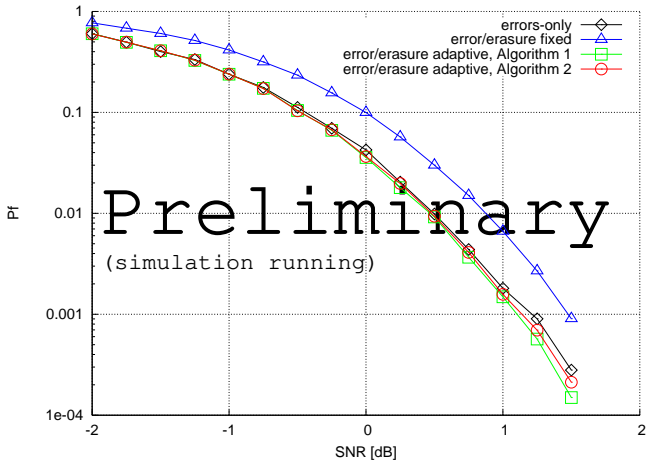}
\caption{Simulation results for $\C(2; 127, 36, 31)$.}
\label{fig:sim127}
\end{figure}

\section{Conclusions}\label{sec:conc}

Despite the seminal results of K\"otter and Vardy about algebraic soft--decision decoding \cite{koetter_vardy:2003} using the {\em Guruswami--Sudan algorithm} \cite{guruswami_sudan:1999}, pseudo--soft decoding with traditional algebraic error/erasure decoders is still of practical interest. Such decoders are widely deployed and efficient implementations are available. Single- and multi--trial error/erasure decoding builds up on these decoders, i.e. they are provided with modified received vectors in which one or multiple sets of most unreliable symbols are erased.

In this paper, we provided two algorithms for adaptive single--trial error/erasure decoding for binary codes. The erasing strategy of the first algorithm is guaranteed to be optimal. The prize for this optimality is computational complexity $\mathcal{O}(n^3)$. The second algorithm gives an approximative optimal solution with precision $10^{-2}$. This allows to reduce complexity to $\mathcal{O}(n^2\sqrt[4]{n})$. Our simulations show that the performance results of both algorithms are virtually indistinguishable in practical settings. However, the approximative algorithm can easily be adapted to fulfill higher precision requirements.

Since our proposed algorithms are optimal, their residual codeword error probability is guaranteed to be superior compared to errors--only decoding and single--trial error/erasure decoding with an optimal fixed erasing strategy. It would be interesting to have an upper bound which proves the gain of adaptive erasing over errors--only and fixed single--trial error/erasure decoding. This bound is in focus of our current investigations.

Our work on the subject is continued with a generalization to multiple decoding trials and non--binary channels. This will enable our algorithms to be applied in existing coding standards which are based on serially concatenated coding schemes.

\section*{Acknowledgments}
\addcontentsline{toc}{section}{Acknowledgments}
The authors would like to thank Serpil Senger and Alexander Zeh for carefully proofreading the manuscript.

\def\noopsort#1{}

\end{document}